\documentclass[onecolumn,draftclsnofoot]{IEEEtran}
\usepackage[utf8]{inputenc}\DeclareUnicodeCharacter{2212}{-} 

\usepackage{amsmath,amsfonts}
\usepackage{algorithmic}
\usepackage{algorithm}
\usepackage{array}
\usepackage[caption=false,font=normalsize,labelfont=sf,textfont=sf]{subfig}
\usepackage{textcomp}
\usepackage{stfloats}
\usepackage{url}
\usepackage{verbatim}
\usepackage{graphicx}
\usepackage{import}
\usepackage[nolist]{acronym}
\usepackage{microtype}
\usepackage{pgf}
\usepackage{pgfplots}
\usepackage{siunitx}
\usepackage[colorinlistoftodos,prependcaption,textsize=small]{todonotes}

\begin{document}
	
\begin{acronym}
	\acro{ads}[ADS]{Advanced Design System}
	\acro{ael}[AEL]{Application Extension Language}
	\acro{af}[AF]{array factor}
	\acro{aod}[AoD]{angle of departure}
	\acro{asic}[ASIC]{application specific integrated circuit}
	\acro{as}[AS]{antenna selection}
	\acro{awgn}[AWGN]{additive white gaussian noise}
	\acro{bs}[BS]{base station}
	\acro{bsl}[BSL]{beam scan loss}
	\acro{cas}[CAS]{computer algebra system}
	\acro{cdf}[CDF]{cumulative distribution function}
	\acro{crc}[CRC]{cyclic redundancy check}
	\acro{csi}[CSI]{channel state information}
	\acro{dac}[DAC]{digital-to-analogue converter}
	\acro{dcm}[LM]{load modulation} 
	\acro{dc}[DC]{direct current}
	\acro{ecb}[ECB]{equivalent complex baseband}
	\acro{embb}[eMBB]{enhanced mobile broadband}
	\acro{enob}[ENOB]{effective number of bits}
	\acro{em}[EM]{electromagnetic}
	\acro{espar}[ESPAR]{electronically steerable passive array radiator}
	\acro{esl}[ESL]{extended system loss}
	\acro{evm}[EVM]{error vector magnitude}
	\acro{fau}[FAU]{Friedrich-Alexander-University Erlangen-Nuremberg}
	\acro{fbs}[FBS]{first bounce scatterer}
	\acro{fc}[FC]{fully connected}
	\acro{fd}[FD]{fully digital}
	\acro{fdd}[FDD]{frequency division duplex}
	\acro{fem}[FEM]{finite element method}
	\acro{fet}[FET]{field-effect transistor}
	\acro{fnbw}[FNBW]{first null beamwidth}
	\acro{fi}[FI]{full illumination}
	\acro{fom}[FOM]{figure of merit}
	\acro{fpga}[FPGA]{field programmable gate array}
	\acro{fr}[FR]{frequency range}
	\acro{fsm}[FSM]{finite state machine}
	\acro{glse}[GLSE]{generalized least-square-error}
	\acro{gscm}[GSCM]{geometry-based stochastic channel model}
	\acro{hadb}[HADB]{hybrid analog-digital beamforming}
	\acro{hpbw}[HPBW]{half power beam width}
	\acro{ic}[IC]{integrated circuit}
	\acro{iid}[i.i.d.]{independent and identically distributed}
	\acro{if}[IF]{intermediate frequency}
	\acro{il}[IL]{insertion loss}
	\acro{isi}[ISI]{inter symbol interference}
	\acro{ima}[IMA]{intermediate amplifier}
	\acro{iot}[IoT]{internet of things}
	\acro{ism}[ISM]{industrial, scientific, and medical}
	\acro{i}[I]{inphase}
	\acro{iq}[IQ]{inphase-quadrature}
	\acro{kpi}[KPI]{key performance indicator}
	\acro{lbs}[LBS]{last bounce scatterer}
	\acro{led}[LED]{light emitting diode}
	\acro{lmmm}[LMMM]{load modulated MIMO}
	\acro{los}[LOS]{line-of-sight}
	\acro{lo}[LO]{local oscillator}
	\acro{lte}[LTE]{Long Term Evolution}
	\acro{mac}[MAC]{medium access control}
	\acro{mad}[MAD]{maximum angular distance}
	\acro{mems}[MEMS]{micro-electro-mechanical systems}
	\acro{mimo}[MIMO]{multiple input multiple output}
	\acro{mmimo}[mMIMO]{massive \ac{mimo}}
	\acro{mmimoabs}[mMIMO]{massive multiple input multiple output}
	\acro{ml}[ML]{maximum likelihood}
	\acro{mmic}[MMIC]{monolithic microwave integrated circuit}
	\acro{mmwave}[mmWave]{millimetre-wave}
	\acro{mse}[MSE]{mean squared error}
	\acro{nlos}[NLOS]{non-\ac{los}}
	\acro{oa}[OA]{outphasing architecture}
	\acro{oe}[OE]{outphasing element}
	\acro{ofdm}[OFDM]{orthogonal frequency-division multiplexing}
	\acro{oma}[OMA]{outphasing MIMO architecture}
	\acro{on}[ON]{outphasing network}
	\acro{oob}[OOB]{out-of-band}
	\acro{opa}[OPA]{outphasing precoder architecture}
	\acro{opex}[OPEX]{operational expenditures}
	\acro{pae}[PAE]{power added efficiency}
	\acro{papr}[PAPR]{peak-to-average power ratio}
	\acro{paspr}[PASPR]{peak-to-average sum-power ratio}
	\acro{pa}[PA]{power amplifier}
	\acro{pcb}[PCB]{printed circuit board}
	\acro{pc}[PC]{partially connected}
	\acro{pdf}[PDF]{probability density function}
	\acro{phy}[PHY]{physical layer}
	\acro{pi}[PI]{partial illumination}
	\acro{pll}[PLL]{phase-locked loop}
	\acro{pm}[PM]{potential minimisation}
	\acro{po}[PO]{power oscillator}
	\acro{psd}[PSD]{power spectral density}
	\acro{pskh}[PSKH]{phase-shift keying on the hypersphere}
	\acro{psk}[PSK]{phase-shift keying}
	\acro{quadriga}[QuaDRiGa]{quasi deterministic radio channel generator}
	\acro{qam}[QAM]{quadrature amplitude modulation}
	\acro{q}[Q]{quadrature}
	\acro{ram}[RAM]{random access memory}
	\acro{ra}[RA]{reflect array}
	\acro{rf}[RF]{radio frequency}
	\acro{rfic}[RFIC]{radio frequency integrated circuit}
	\acro{rl}[RL]{Rotman lens} \acrodefplural{rl}{Rotman lenses}
	\acro{rma}[RMa]{rural macro}
	\acro{rms}[RMS]{root-mean-square}
	\acro{rrc}[RRC]{root-raised-cosine}
	\acro{rx}[RX]{receiver}
	\acro{scm}[SCM]{stochastic channel model}
	\acro{ser}[SER]{symbol error rate}
	\acro{se}[SE]{steering efficiency}
	\acro{sfd}[SFD]{start of frame delimiter}
	\acro{sic}[SIC]{successive interference cancellation}
	\acro{sinr}[SINR]{signal-to-interference-and-noise ratio}
	\acro{siso}[SISO]{single input single output}
	\acro{si}[SI]{separate illumination}
	\acro{sl}[SL]{system loss}
	\acro{sma}[SMA]{subminiature version A}
	\acro{smd}[SMD]{surface mounted device}
	\acro{sm}[SM]{spatial modulation}
	\acro{snr}[SNR]{signal-to-noise ratio}
	\acro{spdt}[SPDT]{single pole, double throw}
	\acro{svd}[SVD]{singular value decomposition}
	\acro{ta}[TA]{transmit array}
	\acro{tdd}[TDD]{time division duplex}
	\acro{trp}[TRP]{total radiated power}
	\acro{ts}[TS]{technical specification}
	\acro{tx}[TX]{transmitter}
	\acro{uma}[UMa]{urban macro}
	\acro{ula}[ULA]{uniform linear array}
	\acro{ura}[URA]{uniform rectangular array}
	\acro{usb}[USB]{universal serial bus}
	\acro{vga}[VGA]{variable gain amplifier}
	\acro{vhdl}[VHDL]{very high speed integrated circuit hardware description language}
	\acro{winner}[WINNER]{wireless world initiative for new radio}
	\acro{xor}[XOR]{exclusive OR}
	\acro{srp}[SRP]{sum of residual powers}
	\acro{zf}[ZF]{zero forcing}
	\acro{rzf}[RZF]{regularized zero forcing}
	\acro{3gpp}[3GPP]{Third Generation Partnership Project}
	\acro{5g}[5G]{fifth generation}
	\acro{6g}[6G]{sixth generation}
\end{acronym}

\title{A High-Level Comparison of Recent\\ Technologies for Massive MIMO Architectures}

\author{Hans Rosenberger, Bernhard Gäde, Ali Bereyhi, Doaa Ahmed, Vahid Jamali,\\ Ralf R. Müller, Georg Fischer, Gaoning He and Mérouane Debbah
\thanks{This work has been submitted to the IEEE for possible publication. Copyright may be transferred without notice, after which this version may no longer be accessible.}%
}

\markboth{}%
{Shell \MakeLowercase{\textit{et al.}}: A Sample Article Using IEEEtran.cls for IEEE Journals}


\maketitle

\begin{abstract}
Since the introduction of \ac{mmimoabs}, the design of a transceiver with feasible complexity has been a challenging problem.
Initially, it was believed that the main issue in this respect is the overall \ac{rf}-cost. 
However, as \ac{mmimoabs} is becoming more and more a key technology for future wireless networks, it is realized, that the \ac{rf}-cost is only one of many implementational challenges and design trade-offs.
In this paper, we present, analyze and compare various novel \ac{mmimoabs} architectures, considering recent emerging technologies such as intelligent surface-assisted and Rotman lens based architectures.
These are compared to the conventional \acl{fd} and \acl{hadb} approaches.
To enable a fair comparison, we account for various hardware imperfections and losses and utilize a novel, universal algorithm for signal precoding.
Based on our thorough investigations, we draw a generic efficiency to quality trade-off for various \ac{mmimoabs} architectures.
We find that in a typical cellular communication setting the reflect/transmit array based architectures sketch the best overall trade-off.
Further, we show that in a qualitative ranking the power efficiency of the considered architectures is independent of the frequency range.
\end{abstract}


\acresetall

\section{Introduction}
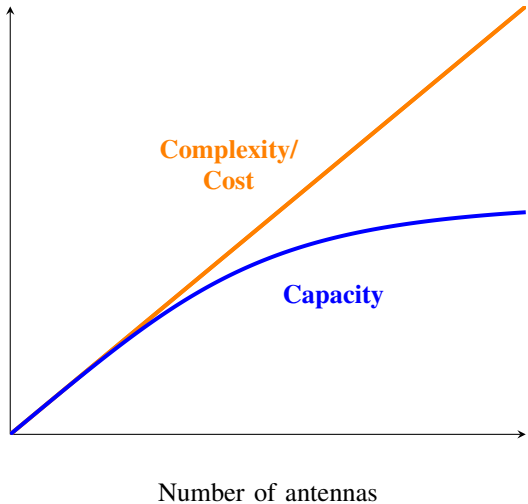
\begin{figure}[!t]
	\centering
	\begin{tikzpicture}
		\begin{axis}[
			axis lines = left,
			xlabel = Number of antennas,
			xtick=\empty,
			ytick = \empty,
			]
			\addplot[
			domain = 0.1:6, 
			samples = 500,
			color = orange,
			line width = 1.5pt,
			]
			{1.5*x} node at (axis cs:2.6,6) {\color{orange} \bfseries Complexity/};
			\addplot[
			domain = 0.1:6, 
			samples = 500,
			color = orange,
			line width = 1.5pt,
			]
			{1.5*x} node at (axis cs:2.6,5.4) {\color{orange} \bfseries Cost};
			]
			\addplot[
			domain = 0.1:6,
			samples = 500,
			color = blue,
			line width = 1.5pt,
			]
			{1.5*x/( (1+ (abs(1.5*x)/5)^3 ) )^(1/3)} node at (axis cs:3.8,3) {\color{blue} \bfseries Capacity};
		\end{axis}
	\end{tikzpicture}
	\caption{Qualitative scaling of the \ac{mmimo} channel capacity versus the \ac{fd} transceiver cost.}
	\label{fig::capacityscaling}
\end{figure}

\IEEEPARstart{F}{or} various wireless communication standards \ac{mimo} has become a crucial technique to enhance throughput and capacity.
This is especially true for mobile communications, where usable spectrum is a valuable commodity. 
For  the \ac{5g} of mobile networks and beyond, even \ac{mmimo} with potentially hundreds or even thousands of antennas at the \ac{bs}, is envisioned to further increase throughput, capacity, coverage, reliability, etc.
However, even for designs with far fewer antennas many challenges arise from an implementation point of view.
Standard approaches such as the \ac{fd} approach, where each antenna element is fed by a dedicated \ac{rf} processing chain, can be uneconomical.
This is mainly due to the fact that the capacity grows linearly with the number of antennas only up to the number of channel-eigenmodes, i.e. the number of non-zero singular values of the channel matrix, afterwards no further multiplexing gain is achieved.
This qualitative scaling is illustrated in Fig.~\ref{fig::capacityscaling}.

This paper tries to provide a fair comparison among some of the most discussed architectures in the literature.
Compared to surveys published before, we aim to compare a wide range of promising architectures under a realistic \ac{rf} system model and performance metrics.
These architectures are briefly introduced in Section~\ref{sec::archs}. Sections~\ref{sec::metrics} and~\ref{sec::meth} introduce common performance metrics and the methodology which is used to establish a fair comparison, respectively.
Section~\ref{sec::sims} presents the numerical investigations. 
Additionally, we expand on the implementational complexity of the architectures in Section~\ref{sec::complexity}.
Lastly, the paper is concluded in Section~\ref{sec::conclusion}.

\section{Architectures}\label{sec::archs}
The \ac{fd} architecture is the most straightforward design.
As the name suggests, each antenna element is equipped with its own dedicated hardware unit, termed an \ac{rf}-chain (see Fig.~2a).
An \ac{rf}-chain consists of two \acp{dac}, an \ac{iq}-modulator, and a unit for conversion to the desired \ac{rf} frequency domain.
This offers the advantage that the transmit signals can be controlled independently at each antenna element.
However, this flexibility comes at the cost of the complexity and the static energy consumption scaling linearly with the number of antennas.
Consequently, the \ac{fd} architecture is not suitable for very large \ac{mimo} systems.

Alternative architectures for \ac{mmimo} can roughly be divided into two groups; namely, \emph{beamforming} and \acfi{dcm} architectures. 
Now, we briefly introduce each group. 
In the subsequent analysis of the architectures, we will then focus primarily on beamforming architectures.

\subsection{Beamforming Architectures}
\begin{figure*}[!t]
	\centering
	\includegraphics[width=\textwidth]{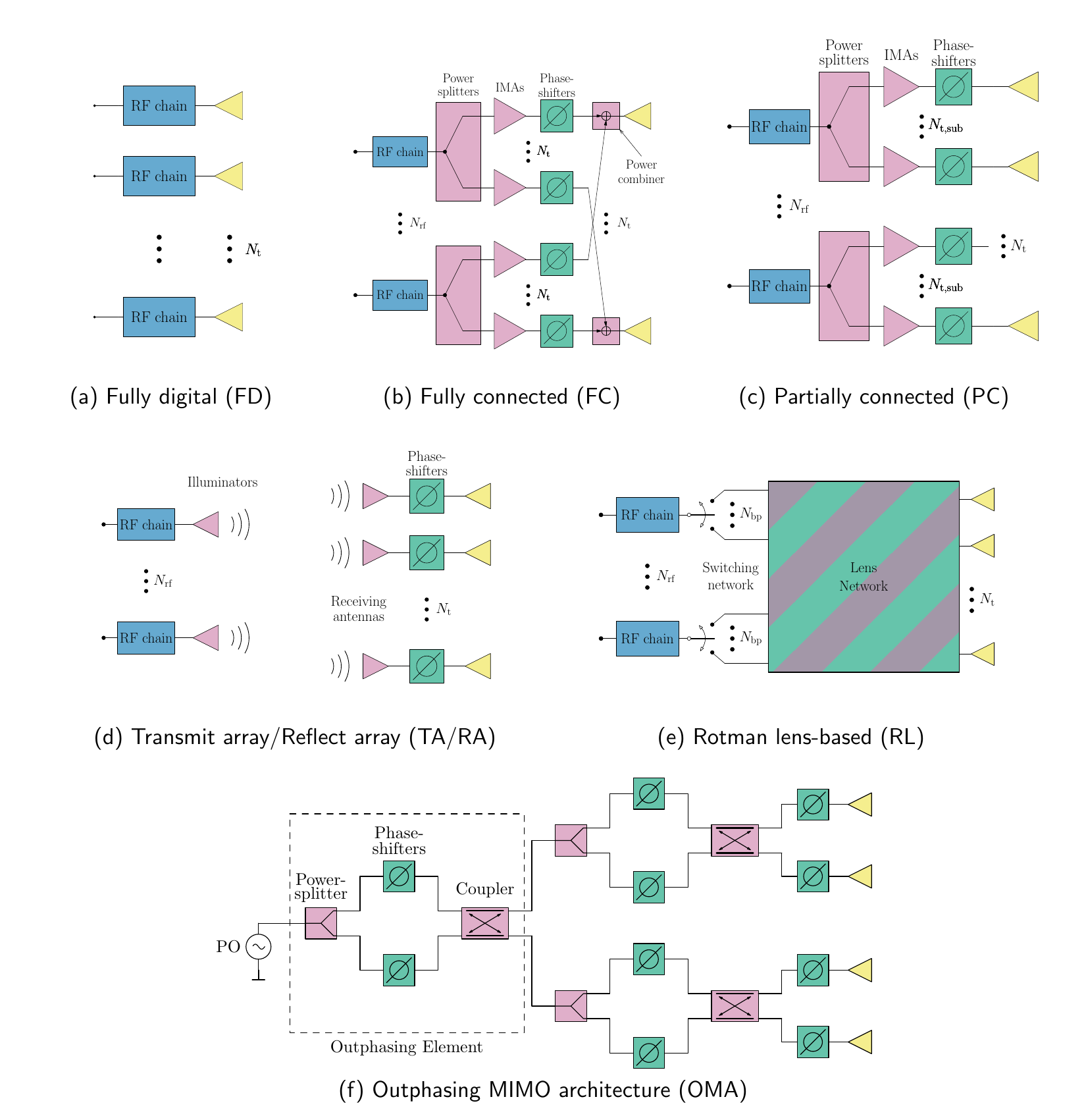}
	\caption{Overview of different architectures for \ac{mmimo} systems.}
	\label{fig::archs}
\end{figure*}

An alternative to the \ac{fd} approach is to reduce the number of \ac{rf}-chains to the number of channel eigenmodes while keeping the number of antennas at the transmitter large.
The \ac{rf}-chains are then connected by a tunable, analogue network to the transmit antennas.
Doing so, the channel dependent beamforming is shifted into the analog domain, while the multiplexing of data streams is still handled by the \ac{rf}-chains.
Throughout the text, we term such architectures \emph{beamforming architectures}.
As the analog network can potentially be implemented by simple and low-cost elements, significant savings in terms of power consumption and system complexity can be achieved.
A schematic overview of these architectures is presented in Fig.~\ref{fig::archs}b-e.

\subsubsection{\Acl{hadb}}
In \ac{hadb} architectures, the analog network is implemented by means of power dividers and combiners to distribute the signal from the \ac{rf}-chains to the antenna elements.
This architecture can be realized in two forms: \Ac{fc} or \ac{pc}.
In the \ac{fc} approach, the output of each \ac{rf}-chain is equally split into $N_\mathrm{t}$ branches to the $N_\mathrm{t}$ antennas. 
Each branch is tuned by a phaseshifter.
The phase-shifted branches with same index are then combined into a power amplifier whose output feeds the corresponding transmit antenna~\cite{Molisch2017Hybrid}.
While this offers nearly the same flexibility as \ac{fd} from the signal processing point of view, the required analog network is extensive.
With a large number of transmit antennas, multiple stages of power dividers are necessary, requiring the compensation of losses by \acp{ima}.
The combiners are typically implemented as Wilkinson combiners, leading to significant dynamic/signal dependent losses, which can severely impact energy efficiency~\cite{Garcia2016Realistic}.
Further, interconnecting the \ac{rf}-chains with all antenna elements (as in \ac{hadb} \ac{fc}) requires a significant amount of crossovers, eg. on a \ac{pcb}, rendering implementation challenging.

A more tractable design is \ac{pc}, in which power combiners are eliminated all together.
Here, each \ac{rf}-chain exclusively feeds only a subset of antennas (see Fig.~2c).
Consequently, the power combiners are no longer necessary, reducing overall losses and system complexity.

\subsubsection{\Acl{ta}/\Acl{ra}}
Realizing the \ac{rf} front-end by a \ac{ta}/\ac{ra}, overcomes the necessity of a large divider (and combiner) network.
Instead, the output of each \ac{rf}-chain feeds to an antenna, referred to as an illuminator, illuminating the aperture of the \ac{ta}/\ac{ra}.
In this approach, the intermediate network is replaced by freespace propagation. 
Beamforming is accomplished by phase-shifts applied at the passive \ac{ta}/\ac{ra} unit~\cite{Reis2019Review}.
The key difference between the \ac{ta} and \ac{ra} is, that for the former the receiving and transmitting array are separate entities, while for the latter the signal is reflected and then radiated again by the same array.
While freespace propagation is an elegant way to feed the antennas, only the superposition of the signals of the \ac{rf} chains can be manipulated at the \ac{ta}/\ac{ra}.
Moreover, different distances between the illuminators and elements of the \ac{ta}/\ac{ra} result in undesirable relative phase and amplitude deviations, affecting the performance.
With \ac{ta}/\ac{ra} based architectures, an illuminator cannot perfectly direct all energy towards the receiving array.
Hence, a part of the electromagnetic wave spills over at the border of the receiving array and is lost. 
To reduce this spillover loss, the directivity of the illuminators can be increased, resulting in a more focused radiation pattern.
Although this reduces the spillover loss, it leads to a more pronounced tapering of the amplitudes across different antenna elements of the \ac{ta}/\ac{ra} unit.
This undesirable tapering effect can lead to a decreased directivity of the array characteristic, termed the taper loss/taper efficiency~\cite{Mailloux2005PhasedHandbook}.
Hence, the geometry of illuminators is to be designed to achieve a trade-off between taper- and spillover losses.

Another design challenge is the coupling between different illuminators.
Compared to \ac{hadb} \ac{fc}, the signals of each \ac{rf}-chain cannot be controlled independently.
Instead, each \ac{ta}/\ac{ra} element receives a superposition of different illuminator signals, depending on the architecture's topology.
To address this design point, we consider two extremes in this study:
In the first setting, each \ac{rf}-chain illuminates the full aperture of the receiving array. 
We refer to this setting as \ac{fi}.
The patterns and orientations of the feed antennas are optimized in this case to achieve a sensible tradeoff between both spillover and taper losses.
Here, some coupling between the different illuminators can be expected, depending on the geometrical arrangement.
In the second extreme case, the receiving array is divided into smaller subarrays, each of which is illuminated by a single feed.
Full decoupling of the illumination of the individual subarrays is achieved either by shielding or simply by spacing them sufficiently far apart.
Each subarray is illuminated by a single dedicated feed, completely decoupling all illuminators~\cite{Jamali2021IRS}.
We refer to this strategy as \ac{si}.

\subsubsection{\Acl{rl} based}
Both \ac{hadb} and the \ac{ta}/\ac{ra} require tuneable components such as phase-shifters. 
However, to achieve a linear phase front (e.g. to focus energy towards a certain \ac{aod}), an electromagnetic lens can be employed.
While there are various ways to implement such lenses, a cost-effective technology is the \ac{rl}.
\Acp{rl} can be fabricated on \acp{pcb} and consist of beam-ports, antenna-ports, and a parallel plate region.
If a beam-port is excited, the resulting signals at the antenna ports are properly delayed, such that constructive interference towards an \ac{aod} is achieved~\cite{Rotman1963Lens}.
As \acp{rl} are planar, two successive stacks are required to steer the beam in two dimensions. 
These stacks form the lens network (see Fig.~2e).
The number of available beamports in the lens network is typically larger than the number of \ac{rf}-chains.
Hence, a switching network is employed to connect the \ac{rf}-chains to the corresponding beamports~\cite{Cho2018rflens}. 

\subsection{\Acl{dcm} Architectures}
The concept of \ac{dcm} takes the idea of beamforming architectures to the extreme.
Here, no \ac{rf}-chains are retained. 
Instead a power oscillator provides the unmodulated carrier signal which is then modulated in the analog domain.
The characteristic difference between load modulation and the beamforming architectures is that in the latter, the analog network needs to be updated on the order of the channel coherence time, while for the former the update rate is required to be greater than the symbol time interval.

Multiple concepts for \ac{dcm}, but also for architectures retaining only a single \ac{rf}-chain~\cite{Mohammadi2012Single}, have been proposed.
A widely discussed technique, is the concept of \ac{espar}, where one active antenna element is surrounded by one or more passive antennas with tuneable loads.
The approach has been validated in~\cite{Alrabadi2011ESPAR}.
Another interesting idea is the application of the outphasing principle to \ac{mmimo}.
Here, the carrier signal, provided by a \ac{po}, is split by the so-called outphasing elements, which can be regarded as tuneable power dividers together with phaseshifting.
Given a fixed transmit power constraint aggregated across all antennas, which is imposed by the output power of the feeding oscillator, arbitrary weights can be realized~\cite{Gaede2021OMA}.

\section{Performance Metrics}\label{sec::metrics}
A characterization of the performance of \ac{mmimo} architectures depends on various factors.
Of primary interest are the energy efficiency and the achievable peak and average throughput.
There are, however, various other soft metrics that play a decisive role, especially for commercial real-world deployment. 
Examples of such metrics are complexity, flexibility, the form factor and upgradeability.
For the sake of brevity, we restrict the metrics discussed in this article to \emph{signal generation accuracy} and \emph{energy efficiency}, which are illustrated in greater detail in the sequel.
Some further notes on the implementational complexity are given in Section~\ref{sec::complexity}.

\subsection{Signal Generation Accuracy}
Since the typical throughput metrics, such as the weighted sum-rate, include various kinds of imperfections, they barely provide insight about the key performance bottlenecks.
More importantly, these metrics often depend on several architecture-independent assumptions such as the channel statistics, the choice of modulation scheme and precoding algorithms.
As a result, such metrics are not universal.
This means that if the assumptions are not met, the characterization becomes generally invalid.

An alternative approach is to characterize the architectures with respect to \textit{signal generation accuracy}, i.e. how accurately an architecture is capable of generating a desired signal.
As a reference scenario, we consider downlink transmission in a multiuser setting.
To ensure fairness, we invoke the \ac{glse} precoding scheme to perform constrained \ac{zf}~\cite{Bereyhi2019GLSE}. 
Unlike typical throughput metrics, the capability of \ac{zf} purely depends on the architecture and hence is a universal measure.
To quantify the \ac{zf} capability of a given architecture, we fix an acceptable average distortion and the transmit power budget across all architectures.
As a distortion measure, we use the \ac{evm}/\ac{mse}. This directly relates to the data rate via the MMSE-mutual information relation on Gaussian channels~\cite{Guo2005MutualInformationMMSE}.
We then compare the aggregated power at the receiver-side averaged over many channel realizations. 
We call this metric in the following the \acfi{sl}.
This metric seems to be natural from a technical viewpoint, there is an interesting intuition behind defining it:
For an architecture with reduced degrees of freedom, the transmitter is more constrained in directing useful signal energy towards the users, given a fixed desired distortion and transmit power.
In this context the ratio of the \ac{sl} between two architectures can also be interpreted as a loss in terms of \ac{sinr}.

\subsection{Power Efficiency}
The \ac{opex} of a \ac{bs} are dominated by the power consumption. 
Hence, power efficiency is a crucial metric.  
Though this metric can be implicitly taken into account via throughput, more detailed insights are obtained by direct characterization.

In general, some part of the energy at the transmitter is not utilized to generate the transmit signal, e.g. the part that is dissipated within an architecture. 
The power efficiency is hence characterized as the fraction of transmit signal power to the total consumed power.
The conventional approach to calculate this fraction is to set the transmit power to the \ac{trp}, i.e. the power radiated by the transmitter's front-end.
However, doing so, we neglect a key aspect of an architecture, namely, to direct energy in space, i.e., to increase the power flux density at the receivers.
To take this into account, we use the following approach for characterization of the power efficiency:
As a benchmark, we define an ideal, lossless beamforming architecture that can perform arbitrary linear beamforming at its front-end antenna array, given a fixed total transmit power.
All architectures are then evaluated against this ideal architecture in a single-user \ac{los} scenario, where the user is located at a fixed direction relative to the transmitter.
For this scenario, we optimize each architecture such that it maximizes the received power density at the user, i.e. it steers the main beam towards the user.
We then quantify the capability of each architecture by determining the \acfi{se}:
The \ac{se} determines the loss in directivity towards the direction of the user, relative to the ideal reference architecture, while also considering power dissipation.

\begin{figure*}[!t]
	\centering
	\includegraphics[width=\textwidth]{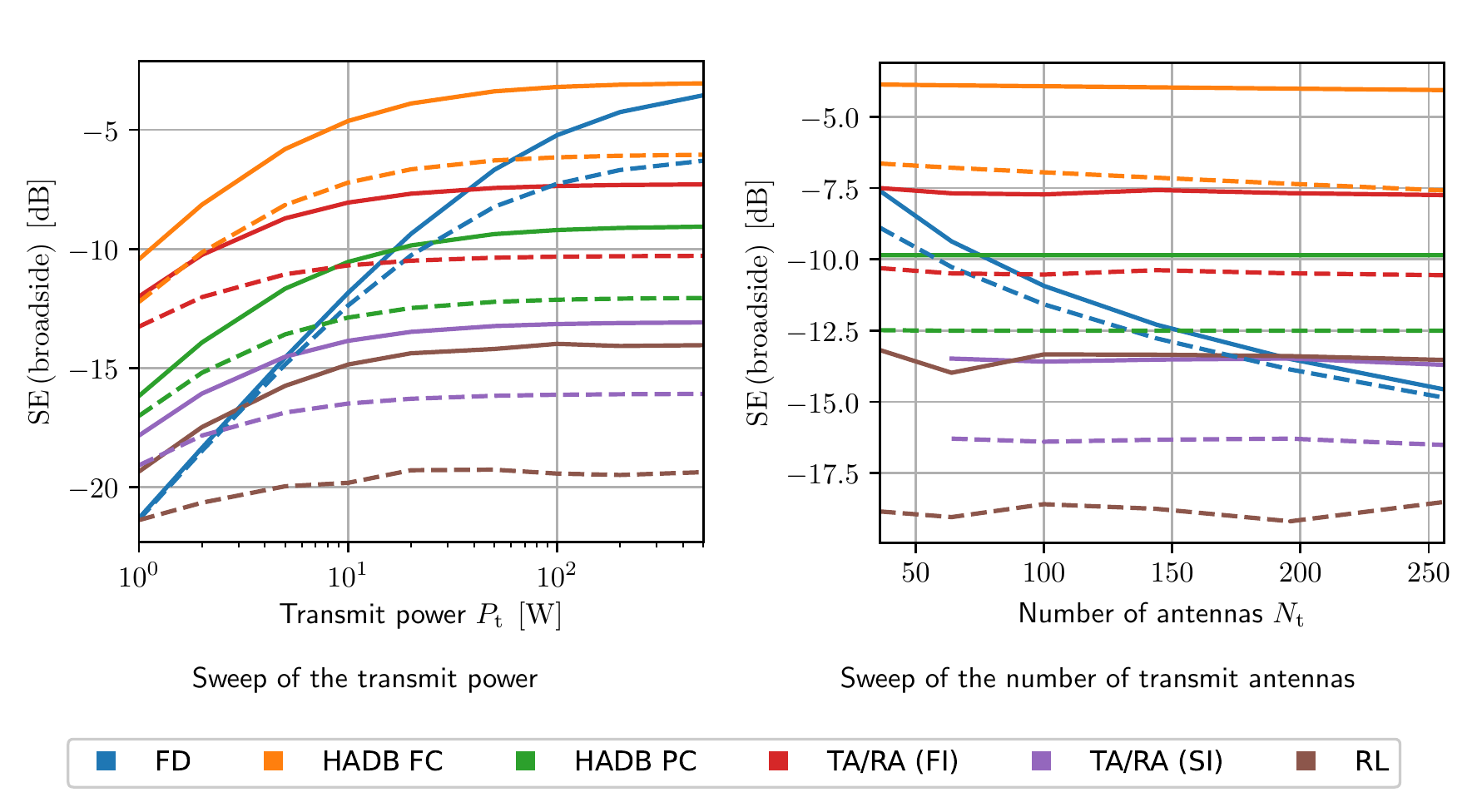}
	\caption{Sweeps of the steering efficiency for a design carrier frequency of \SI{3.5}{\giga\hertz} (solid lines) and \SI{28}{\giga\hertz} (dashed lines)}
	\label{fig::steeringeff}
\end{figure*}

\section{Methodology}\label{sec::meth}
Now, we expand on the methodology and the system modeling used in the subsequent numerical evaluations.
For the sake of brevity, we only restrict our attention to the most important aspects of the system model.
The relevant default parameter set is summarized in Table~\ref{tab::defaultparams}.

\begin{table}[!t]
	\centering
	\caption{Reference Simulation Parameters}
	\label{tab::defaultparams}
	\begin{tabular}{|c|c|}
		\hline
		Transmit power $P_\mathrm{t}$ & \SI{20}{\watt}\\
		\hline
		Number of \ac{rf}-chains/simultaneous users & 4 \\ 
		\hline
		Power amplifier \acl{pae} $\eta_{\mathrm{PA}}$ & \SI{50}{\percent} \\
		\hline
		Loss per phase-shifter & \SI{1}{\decibel}\\
		\hline
		Phase-shifter resolution & \SI{2}{bit}\\
		\hline
		\ac{rf}-chain power consumption & \SI{2.08}{\watt}\\
		\hline
		Design frequency $f = c_0/\lambda$ & \SI{3.5}{\giga\hertz}\\
		\hline
		Number of antennas $N_\mathrm{t}$ (Square URA) & $64$\\
		\hline
		Antenna element spacing 
		& $\lambda/2$\\
		\hline
	\end{tabular}
\end{table}

The main design frequency is chosen within \ac{fr} 1, covering the sub-\SI{6}{\giga\hertz} domain.
This choice of frequency is derived from the following considerations:
While there is significant potential for \ac{mmimo} in the mm-Wave range, termed \ac{fr} 2, there is still a great interest on the more commercial frequency ranges as wider and more reliable coverage is achieved. 
Nonetheless, we still provide some simulations for the energy efficiency in the \ac{mmwave} range at \SI{28}{\giga\hertz}, too.
For realistic channel modeling we employ the 3D~\ac{quadriga}~\cite{Jaeckel2014Quadriga}, which is compliant to current \ac{3gpp} specifications.

Since the consumption of a single \ac{rf}-chain plays a decisive role in power consumption, we consider a conservative value of \SI{2.08}{\watt}. 
Which is easily achieved via an \ac{rf}-chain assembled by discrete components.
Note that much lower values, e.g. \SI{200}{\milli\watt}, are achievable through a combined on-chip integration of \ac{rf}-chains as reported in~\cite{Lee20195GNovelTransceiver}.
Unless otherwise stated, we assume an $8 \times 8$ \ac{ura} whose patch antenna elements are spaced $\lambda/2$ apart.
The radiation patterns of these elements are determined by a full \ac{em} simulation and integrated into the channel model.
Further, for the individual \acp{rl} an \ac{em} simulation of a design with five beamports and eight antenna ports is employed.
Phase-shifters are assumed to be quantized with a resolution of \SI{2}{bit} and a loss of \SI{1}{\decibel}.
For other \ac{rf} components not explicitly discussed here, we resort to values obtained by simulations and an extensive literature research.

\section{Numerical Evaluations}\label{sec::sims}
Next, we investigate the architectures with respect to the proposed metrics, where we use the methodology laid out in the previous section.

\begin{table*}[!t]
	\centering
	\caption{\acl{sl}: Analysis of Signal Generation Accuracy}
	\label{tab::sl}
	\begin{tabular}{|c|c|c|c|c|c|c|}
		\hline
		Scenario & \acs{fd} & \acs{hadb} \acs{fc} & \acs{hadb} \acs{pc} & \ac{ta}/\acs{ra} (\acs{fi}) & \acs{ta}/\acs{ra} (\acs{si}) & \acs{rl} \\
		\hline
		\acs{uma} \acs{los} $\acs{sl}_\mathrm{rel}$ & \SI{0.0}{\decibel} & \SI{-2.7}{\decibel} & \SI{-8.2}{\decibel} & \SI{-7.3}{\decibel} & \SI{-8.3}{\decibel} & \SI{-20.2}{\decibel} \\
		\hline
		\acs{uma} \acs{nlos} $\acs{sl}_\mathrm{rel}$ & \SI{0.0}{\decibel} & \SI{-2.7}{\decibel} & \SI{-8.1}{\decibel} & \SI{-7.3}{\decibel} & \SI{-8.5}{\decibel} & \SI{-19.2}{\decibel} \\
		\hline
		\acs{rma} \acs{los} $\acs{sl}_\mathrm{rel}$ & \SI{0.0}{\decibel} & \SI{-2.7}{\decibel} & \SI{-8.2}{\decibel} & \SI{-7.4}{\decibel} & \SI{-8.3}{\decibel} & \SI{-20.0}{\decibel} \\
		\hline
		\acs{rma} \acs{nlos} $\acs{sl}_\mathrm{rel}$ & \SI{0.0}{\decibel} & \SI{-2.8}{\decibel} & \SI{-8.2}{\decibel} & \SI{-7.5}{\decibel} & \SI{-8.4}{\decibel} & \SI{-19.8}{\decibel} \\
		\hline
	\end{tabular}
\end{table*}

As the first experiment, we analyze the performance of the beamforming architectures with respect to signal generation accuracy.
The simulations are conducted for different channel scenarios, representing \ac{uma} and \ac{rma} environments as specified in~\cite{3GPP38901}.
Both scenarios are simulated for \ac{los} and \ac{nlos} conditions separately.
The results are compiled in Table~\ref{tab::sl}.
The values are normalized to the performance of the \ac{fd} reference architecture, as this architecture has the highest degrees of freedom across all other architectures.
Performing slightly worse than \ac{fd} is the \ac{hadb} \ac{fc} architecture.
Although each \ac{rf}-chain is connected to all antenna elements, the quantized phase-shifts and the lack of amplitude control lead to a performance degradation.\footnote{Theoretically, it has been shown that \ac{hadb} \ac{fc} can achieve equal performance to \ac{fd}~\cite{Molisch2017Hybrid}.}
With some distance both the \ac{ta}/\ac{ra} and the \ac{hadb} \ac{pc} architectures follow the \ac{hadb} \ac{fc}.
Both, \ac{hadb} \ac{pc} and \ac{ta}/\ac{ra} (\ac{si}) perform very similar in all scenarios. 
This is intuitive, since the only difference between these two architectures are the static transfer weights between the \ac{rf}-chains and the connected subsets of antenna elements.
The \ac{ta}/\ac{ra} (\ac{fi}) performs slightly better than its separately illuminated counterpart.
Note that the four illuminators were arranged in a way to minimize mutual coupling.
Lastly with considerable distance (more than \SI{10}{\decibel}) the \ac{rl}--based architecture performs worst.
The design of this architecture offers the precoder a set of only 25 predefined beams to choose from, leading to a much lower flexibility in terms of signal generation accuracy.
While increasing the number of beamports would improve the \ac{rl}--based architecture in terms of signal generation accuracy, it would in turn degrade the energy efficiency.

To investigate the power efficiency, the \ac{se} is plotted against the transmit power and the number of antennas at the front-end for different beamforming architectures in Fig.~\ref{fig::steeringeff}.
For both frequency ranges, the numerical results show the same behaviour, differing only by a shift along the ordinate axis.
This is due to increased losses in transmission lines and components, such as \ac{rf}-switches.
These may also affect different architectures to a different degree, however, the effect does not impact the ranking of the architectures.
With respect to the transmit power, the \ac{fd} architecture shows the steepest decline in power efficiency towards small transmit powers.
This is due to the static \ac{dc} power consumption of the \ac{rf}-chains.
In fact, for low transmit powers, the static power consumption dominates the overall energy consumption.
A similar decline is observed for the other architectures, albeit to a smaller extend. 
The reduced dependence on the transmit power is explained by the lower number of \ac{rf}-chains and thus a considerably lower static \ac{dc} consumption.
Interestingly, all other beamforming architectures (except \ac{hadb} \ac{fc}) saturate at a much lower energy efficiency for high transmit powers.
This behaviour is mainly caused by the passive network which follows the \acp{pa}.
For example, in the \ac{ta}/\ac{ra} the \acp{pa} are integrated into the illuminators, such that they have to compensate for propagation losses and the passive array to achieve a given transmit power.
At high transmit powers, this loss becomes significant.
For the \ac{fd} and \ac{hadb} architectures this loss can be generally avoided by placing the \acp{pa} directly in front of the antenna elements. 

Considering the sweep of the number of transmit antennas, a pronounced drop in energy efficiency is observed for \ac{fd} transmission.
This is as the number of \ac{rf}-chains scales with the number of antennas.
For the other beamforming architectures the number stays constant, but a slope is observed for \ac{hadb} \ac{fc}.
In this respect, the \acp{ima} are the deciding factor:
As the number of dividers and combiners grows large, more amplifiers are necessary to compensate for losses.
Noting, that the slope is more pronounced for the \ac{mmwave} range, as in this range the power consumption of the amplifiers is larger.
For the \ac{ta}/\ac{ra}-based architecture the energy efficiency stays approximately constant when the number of antennas is scaled up, since no additional active components are required.
The latter behavior describes a clear advantage of the \ac{ta}/\ac{ra}-based architecture.

\section{Complexity}\label{sec::complexity}
Compared to \ac{fd}, all alternative architectures significantly reduce the complexity of the \ac{rf} generation network.
However, this reduction is not for free, as complexity is shifted to the analog network.
To see this, consider the following example: From Section~\ref{sec::sims}, we know that \ac{hadb} \ac{fc} can be designed such to meet the signal generation accuracy of \ac{fd}.
However, this comes at a high complexity of the analog network.
For a system with 256 antennas and four \ac{rf}-chains, this leads to more than thousand connections/transmission lines, power dividers and combiners.
This design task is commercially challenging, since a high number of crossovers between lines is required. 
The network complexity is greatly reduced in \ac{hadb} \ac{pc}, or even more efficiently in the \ac{ta}/\ac{ra}-based architecture.
In the latter architecture, the divider network is essentially for free, as signal splitting is accomplished over the air.
This renders this architecture suitable for the scaling desired in \ac{mmimo} transmitters.
As observed in Section~\ref{sec::sims}, these complexity reductions come at the cost of reduced flexibility in signal generation accuracy. 

\begin{table}[!t]
	\centering
	\caption{Comparison of different beamforming architectures with respect to relevant performance metrics.}
	\label{tab::thetable}
	\begin{tabular}{|c|c|c|c|c|c|}
		\hline
		& \ac{fd} & \ac{fc} & \ac{pc} & \ac{ta}/\ac{ra} & \ac{rl}-based \\
		\hline
		Power Efficiency & $\circ$ & + & $\circ$ & + & -- \\
		\hline
		Signal Generation Accuracy & + & + & $\circ$ & $\circ$ & --  \\
		\hline
		Complexity & -- & -- & + & + & $\circ$ \\
		\hline
	\end{tabular}
\end{table}

\section{Conclusion}\label{sec::conclusion}
\ac{mmimo} is a versatile technology and a single best architecture is nearly impossible to identify for all use cases.
As we discussed in this paper, the impact of the desired average transmit power, the number of antennas as well as the system complexity play a crucial role in ranking the architectures.
Ultimately, these choices depend on the specific needs of an operator and the use-case in mind.

The trend in \ac{bs} technology progresses towards a less centralized approach with a larger number of \acp{bs} with a smaller coverage area and thus a lower transmit power.
For such a commercial use case, the \ac{ta}/\ac{ra} based architecture offers a fair complexity-performance trade-off for \ac{mmimo} transmission.
A high-level ranking of the overall performance in this regime is given in Table~\ref{tab::thetable}.

\bibliographystyle{IEEEtran}
\bibliography{bibliography}

\begin{thebibliography}{10}
\providecommand{\url}[1]{#1}
\csname url@samestyle\endcsname
\providecommand{\newblock}{\relax}
\providecommand{\bibinfo}[2]{#2}
\providecommand{\BIBentrySTDinterwordspacing}{\spaceskip=0pt\relax}
\providecommand{\BIBentryALTinterwordstretchfactor}{4}
\providecommand{\BIBentryALTinterwordspacing}{\spaceskip=\fontdimen2\font plus
\BIBentryALTinterwordstretchfactor\fontdimen3\font minus
  \fontdimen4\font\relax}
\providecommand{\BIBforeignlanguage}[2]{{%
\expandafter\ifx\csname l@#1\endcsname\relax
\typeout{** WARNING: IEEEtran.bst: No hyphenation pattern has been}%
\typeout{** loaded for the language `#1'. Using the pattern for}%
\typeout{** the default language instead.}%
\else
\language=\csname l@#1\endcsname
\fi
#2}}
\providecommand{\BIBdecl}{\relax}
\BIBdecl

\bibitem{Molisch2017Hybrid}
A.~F. Molisch, V.~V. Ratnam, S.~Han, Z.~Li, S.~L.~H. Nguyen, L.~Li, and
  K.~Haneda, ``Hybrid beamforming for massive {MIMO}: A survey,'' \emph{IEEE
  Communications Magazine}, vol.~55, no.~9, pp. 134--141, 2017.

\bibitem{Garcia2016Realistic}
A.~Garcia-Rodriguez, V.~Venkateswaran, P.~Rulikowski, and C.~Masouros, ``Hybrid
  analog–digital precoding revisited under realistic {RF} modeling,''
  \emph{IEEE Wireless Communications Letters}, vol.~5, no.~5, pp. 528--531,
  2016.

\bibitem{Reis2019Review}
J.~R. Reis, M.~Vala, and R.~F.~S. Caldeirinha, ``Review paper on transmitarray
  antennas,'' \emph{IEEE Access}, vol.~7, pp. 94\,171--94\,188, 2019.

\bibitem{Mailloux2005PhasedHandbook}
R.~Mailloux, \emph{Phased Array Antenna Handbook}, 2nd~ed.\hskip 1em plus 0.5em
  minus 0.4em\relax Artech House, 2005.

\bibitem{Jamali2021IRS}
V.~Jamali, A.~M. Tulino, G.~Fischer, R.~R. Müller, and R.~Schober,
  ``Intelligent surface-aided transmitter architectures for millimeter-wave
  ultra massive {MIMO} systems,'' \emph{IEEE Open Journal of the Communications
  Society}, vol.~2, pp. 144--167, 2021.

\bibitem{Rotman1963Lens}
W.~Rotman and R.~Turner, ``Wide-angle microwave lens for line source
  applications,'' \emph{IEEE Transactions on Antennas and Propagation},
  vol.~11, no.~6, pp. 623--632, 1963.

\bibitem{Cho2018rflens}
Y.~J. Cho, G.-Y. Suk, B.~Kim, D.~K. Kim, and C.-B. Chae, ``{RF} lens-embedded
  antenna array for mm{W}ave {MIMO}: Design and performance,'' \emph{IEEE
  Communications Magazine}, vol.~56, no.~7, pp. 42--48, 2018.

\bibitem{Mohammadi2012Single}
A.~Mohammadi and F.~M. Ghannouchi, ``Single {RF} front-end {MIMO}
  transceivers,'' \emph{IEEE Communications Magazine}, vol.~49, no.~12, pp.
  104--109, 2011.

\bibitem{Alrabadi2011ESPAR}
O.~N. Alrabadi, C.~Divarathne, P.~Tragas, A.~Kalis, N.~Marchetti, C.~B.
  Papadias, and R.~Prasad, ``Spatial multiplexing with a single radio:
  Proof-of-concept experiments in an indoor environment with a 2.6-{GHz}
  prototype,'' \emph{IEEE Communications Letters}, vol.~15, no.~2, pp.
  178--180, 2011.

\bibitem{Gaede2021OMA}
B.~Gäde, S.~Erhardt, G.~Fischer, and R.~R. Müller, ``An outphasing {MIMO}
  architecture prototype,'' in \emph{2020 50th European Microwave Conference
  (EuMC)}, 2021, pp. 208--211.

\bibitem{Bereyhi2019GLSE}
A.~Bereyhi, M.~A. Sedaghat, R.~R. Müller, and G.~Fischer, ``{GLSE} precoders
  for massive {MIMO} systems: Analysis and applications,'' \emph{IEEE
  Transactions on Wireless Communications}, vol.~18, no.~9, pp. 4450--4465,
  2019.

\bibitem{Guo2005MutualInformationMMSE}
D.~Guo, S.~Shamai, and S.~Verdu, ``Mutual information and minimum mean-square
  error in {Gaussian} channels,'' \emph{IEEE Transactions on Information
  Theory}, vol.~51, no.~4, pp. 1261--1282, 2005.

\bibitem{Jaeckel2014Quadriga}
S.~Jaeckel, L.~Raschkowski, K.~Börner, and L.~Thiele, ``{QuaDRiGa}: A {3-D}
  multi-cell channel model with time evolution for enabling virtual field
  trials,'' \emph{IEEE Transactions on Antennas and Propagation}, vol.~62,
  no.~6, pp. 3242--3256, 2014.

\bibitem{Lee20195GNovelTransceiver}
J.~Lee, S.~Han, J.~Lee, B.~Kang, J.~Bae, J.~Jang, S.~Oh, J.-S. Chang, S.~Kang,
  K.~Y. Son, H.~Lim, D.~Jeong, I.~Jong, S.~Baek, J.~H. Lee, R.~Ni, Y.~Zuo,
  C.-W. Yao, S.~Heo, T.~B. Cho, and I.~Kang, ``A {Sub-6-GHz} {5G} {New Radio}
  {RF} transceiver supporting {EN-DC} with {3.15-Gb/s} {DL} and {1.27-Gb/s}
  {UL} in 14-nm {FinFET} {CMOS},'' \emph{IEEE Journal of Solid-State Circuits},
  vol.~54, no.~12, pp. 3541--3552, 2019.

\bibitem{3GPP38901}
\BIBentryALTinterwordspacing
3GPP, ``{Study on channel model for frequencies from 0.5 to 100 GHz},'' {3rd
  Generation Partnership Project (3GPP)}, Technical Specification (TS) 38.901,
  01 2020, version 16.1.0. [Online]. Available:
  \url{https://portal.3gpp.org/desktopmodules/Specifications/SpecificationDetails.aspx?specificationId=3173}
\BIBentrySTDinterwordspacing

\end{thebibliography}

\vfill

\end{document}